\documentclass[a4paper,fleqn,usenatbib,useAMS]{mnras}

\usepackage[T1]{fontenc}
\usepackage{ae,aecompl}

\usepackage{graphicx}	
\usepackage{amsmath}	
\usepackage{amssymb}	

\usepackage{times}
\usepackage{epstopdf}
\usepackage{txfonts}
\usepackage{natbib}
\usepackage{array}
\usepackage{multirow} 	
\usepackage{pdflscape}	
\usepackage{booktabs}

\usepackage{chngcntr} 
\counterwithout*{figure}{section}
\counterwithout*{table}{section}

\title[global VLBI imaging of MG~J0751+2716]{SHARP -- V.~ Modelling gravitationally-lensed radio arcs imaged with global VLBI observations}

\author[C. Spingola et al.]{C. Spingola,$^{1}$\thanks{E-mail: spingola@astro.rug.nl} J. P. McKean,$^{1,2}$ M. W. Auger,$^3$ C. D. Fassnacht,$^4$ L. V. E. Koopmans,$^{1}$
\newauthor D. J. Lagattuta$^5$ and S. Vegetti$^{6}$
\\
$^1$Kapteyn Astronomical Institute, University of Groningen, Postbus 800, NL-9700 AV Groningen, the Netherlands\\
$^2$ASTRON, Netherlands Institute for Radio Astronomy, Oude Hoogeveensedijk 4, 7991 PD Dwingeloo, the Netherlands\\
$^3$Institute of Astronomy, University of Cambridge, Madingley Rd, Cambridge CB3 0HA, United Kingdom\\
$^4$Department of Physics, University of California, Davis, CA 95616, USA\\
$^5$Univ Lyon, Univ Lyon1, Ens de Lyon, CNRS, Centre de Recherche Astrophysique de Lyon UMR5574, F-69230 Saint-Genis-Laval, France\\
$^6$Max Planck Institute for Astrophysics, Karl-Schwarzschild-Strasse 1, 85740 Garching, Germany
}

\date{Accepted. Received; in original form 2017 December 14}

\pubyear{2017}

\begin{document}
\label{firstpage}
\pagerange{\pageref{firstpage}--\pageref{lastpage}}
\maketitle


\begin{abstract}
We present milliarcsecond (mas) angular resolution observations of the gravitationally lensed radio source MG J0751+2716 (at $z=3.2$) obtained with global Very Long Baseline Interferometry (VLBI) at 1.65 GHz. The background object is highly resolved in the tangential and radial directions, showing evidence of both compact and extended structure across several gravitational arcs that are 200 to 600~mas in size. By identifying compact sub-components in the multiple images, we constrain the mass distribution of the foreground $z=0.35$ gravitational lens using analytic models for the main deflector [power-law elliptical mass model; $\rho(r) \propto r^{-\gamma}$, where $\gamma=2$ corresponds to isothermal] and for the members of the galaxy group. Moreover, our mass models with and without the group find an inner mass-density slope steeper than isothermal for the main lensing galaxy, with $\gamma_1 = 2.08 \pm 0.02$ and $\gamma_2 = 2.16 \pm 0.02$ at the 4.2$\sigma$ level and 6.8$\sigma$ level, respectively, at the Einstein radius ($b_1 = 0.4025 \pm 0.0008$ and $b_2 = 0.307 \pm 0.002$ arcsec, respectively). We find randomly distributed image position residuals of about 3 mas, which are much larger that the measurement errors ($40$ $\mu$as on average). This suggests that at the mas level, the assumption of a smooth mass distribution fails, requiring additional structure in the model. However, given the environment of the lensing galaxy,  it is not clear whether this extra mass is in the form of sub-haloes within the lens or along the line of sight, or from a more complex halo for the galaxy group. 
\end{abstract}



\section{Introduction}

In observational cosmology, gravitational lensing is the only method that allows one to directly probe the projected matter (including dark matter) density distribution of galaxies over an extended range of scales, independent of its dynamical state (see \citealt{Treu2010} for a recent review). Ever since the first discovery of multiple imaging \citep*{Walsh1979}, gravitational lensing has been used to investigate a broad range of astrophysical questions, from the structure of the large-scale matter distribution to the physical properties of the individual lensing galaxies \citep{Koopmans2009, Giocoli2013, Sonnenfeld2015}. In particular, gravitational lensing is a powerful technique to test models for global halo profiles that are predicted from hierarchical galaxy formation simulations (e.g. \citealt*{Navarro1997}). However, the constraints provided by observations of the multiple images alone are often not sufficient to determine an univocal lens mass model. It is for this reason that parametric models, motivated by the observed general properties of typical galaxies in the local Universe, are generally used to overcome this obstacle \citep{Keeton2001a,Wucknitz2001}.  

The simplest macro model that can reproduce the relative positions and flux-ratios of the multiple images is the singular isothermal ellipsoid (SIE) plus an external shear to account for neighbouring galaxies, which is often sufficient to describe the projected mass of elliptical lensing galaxies. For example, the majority of the lenses in the optical  CfA-Arizona-Space Telescope Lens Survey (CASTLES) can be modelled by simple ellipsoidal mass distributions with an external shear, and the number of gravitational lenses that require a deviation from this model are few \citep{Munoz1998, Falco1999}. Among all of the known galaxy-scale gravitational lenses, those with extended images of the background lensed source provide the most constraints to test the mass model, and, therefore, they can be used to investigate whether an elliptical power-law density model is a more accurate description of the matter content of the lensing galaxy (e.g. \citealt{Koopmans2009}). Also, from comparing the constraints derived for data at progressively higher angular resolution, \citet{Lagattuta2012} find that the parameter space is highly constrained with better quality imaging.

In this respect, interferometric observations at radio wavelengths can currently provide the highest angular resolution imaging available, and surveys of radio-loud gravitational lens systems at high angular resolution have been carried out extensively \citep{Burke1990, Hewitt1992, Patnaik1993, Winn2001, Myers2003, Browne2003}. However, most of the lensed sources discovered in this way have compact structure and the unresolved multiple images place only a few constraints on the lens mass model.  This is because many of these objects were discovered through a systematic search of flat-spectrum radio sources, which are typically compact when observed with the Very Large Array (VLA at 8.46~GHz; 170~mas beam size) and the Multi-Element Radio Linked Interferometry Network (MERLIN at 5~GHz; 50~mas beam size). In a few cases, the radio structure of the background source is resolved into Einstein rings and extended gravitational arcs (e.g. \citealt{Hewitt1988,Langston1989,Biggs2001}), as for example, the lenses MG~J0414+0534 \citep{MacLeod2013}, MG~B2016+112 \citep{Koopmans2002, More2009}, JVAS~B1933+503 \citep{Suyu2012} and CLASS B1555+375 \citep{Hsueh2016}. In these cases, the extra constraints provided by the extended arcs revealed that the macro model could not be explained by a simple SIE, but additional mass structures are necessary to reproduce the relative position and fluxes of the observed images.  

Gravitational lensing is a powerful method to directly detect and quantify any deviation of a smooth mass model for the primary lens and it can be used with this aim in two different ways. One method uses the flux ratios of the multiple images of compact background sources to find evidence for peculiar magnifications, which can be due to a perturbation from small-scale structures in the lensing galaxy \citep{Mao1998}. For example, the study of seven radio-loud flux-ratio anomalous lenses by \citet{DalalKochanek2002} finds that the mass fraction of substructure required to reproduce the flux ratios within their sample was $ f_{\rm sub}$ = 2$^{+5.0}_{-1.6}$  percent (90 percent confidence levels). The other method to detect perturbations to a smooth mass model consists of observing astrometric anomalies of the multiple images, namely observing lensed images in different positions from what is expected from a smooth mass distribution. Given the high angular resolution from Very Long Baseline Interferometry (VLBI) data at cm-wavelengths (2--10~mas beam size), deviations of a smooth macro model can be found via astrometric perturbations of the multiple images; the additional mass structure can perturb the deflection angle and, therefore, the relative positions of the images can be shifted. For example, simulations predict that a dark matter sub-halo of a mass $10^8$~M$_{\odot}$ in a Milky Way-sized galaxy can produce an astrometric perturbation of $\sim10$ mas in the lensed images \citep{MetcalfMadau2001}. Even if the sub-halo has a lower mass ($>10^{5}$~M$_{\odot}$), it is still possible to observe local independent distortions in the VLBI images of lensed radio jets \citep{Metcalf2002}. 
These distortions, which appear as bends in the jets, can be detected by measuring the local curvature of extended multiple images and noting differences between points that correspond to the same source position. Therefore, observations with mas (and sub-mas)  angular resolution are key to testing the smoothness of the macro models.

With these aims, the Strong lensing at High Angular Resolution Program (SHARP; Fassnacht et al., in prep) has carried out high angular resolution observations at optical, near-infrared (NIR) and radio wavelengths of known gravitational lenses with extended source structure. The main goal is to detect and measure possible perturbations to the macro models associated with the main lensing galaxy halo to constrain models for galaxy formation, dark matter and cosmology. Thus far, SHARP has focussed on using high angular resolution observations with the adaptive optics system on the W. M. Keck 10-m Telescope and the {\it Hubble Space Telescope} ({\it HST}) to test global mass models \citep{Lagattuta2010,Lagattuta2012,Chen2016}, infer the properties of luminous and dark dwarf galaxies (substructures; \citealt{McKean2007,Vegetti2012}) and investigate the mass perturbations due to galactic-scale disks \citep{Hsueh2016,Hsueh2017}. Here, we extend SHARP to higher angular resolution (by a factor of 30--60) and longer wavelengths by presenting imaging of the most extended gravitational arc on mas-scales known with VLBI at cm-wavelengths.

The first target is the strongly lensed radio-loud quasar MG~J0751+2716, which is one of the most promising targets for studying the smoothness of the lensing mass distribution on mas-scales. The gravitational lens was discovered as part of the VLA follow-up of sources found from the MIT--Green Bank survey, and shows large gravitational arcs at high surface brightness when observed with MERLIN at 5 GHz (50~mas beam size; \citealt{Lehar1997}). Optical imaging shows that the lensing galaxy is a satellite of a bright cluster galaxy (BCG) and is part of a small group of galaxies. The redshift of this group of galaxies was found to be $z_{\rm group} = 0.35$ \citep{Tonry1998, Momcheva2006} and the redshift of the background source is $z = 3.200$ \citep{Tonry1998,Alloin2007}.  The macro models proposed to date require a significant external shear, likely due to the group of galaxies in the field \citep{Lehar1997, Momcheva2006, Alloin2007}. The large extent of the arcs coupled with the bright emission of this source ($\sim 350$ mJy at 1.7 GHz) make MG~J0751+2716 an excellent lens system to study the level of deviations from a smooth macro model.

In this paper, we present new global VLBI observations of MG J0751+2716 at 1.65 GHz with the main aim of investigating the global mass model and determining a precise radial density profile of the mass distribution (accounting for the galaxy group). Our paper is organized as follows. In Section \ref{obs} we describe the observations and imaging results, and in Section \ref{model} we illustrate the improved macro models that can be determined with mas-scale angular resolution imaging. Our discussion and summary of results are presented in Sections \ref{disc} and \ref{conc}, respectively. Throughout this paper, we assume $H_0=67.8\; \mathrm{km\,s^{-1}\; Mpc^{-1}}$, $\Omega_{\rm M}=0.31$, $\Omega_{\Lambda}=0.69$ in a flat Universe \citep{Planck2015}.


\begin{figure*}
\centering
\includegraphics[scale = 0.85]{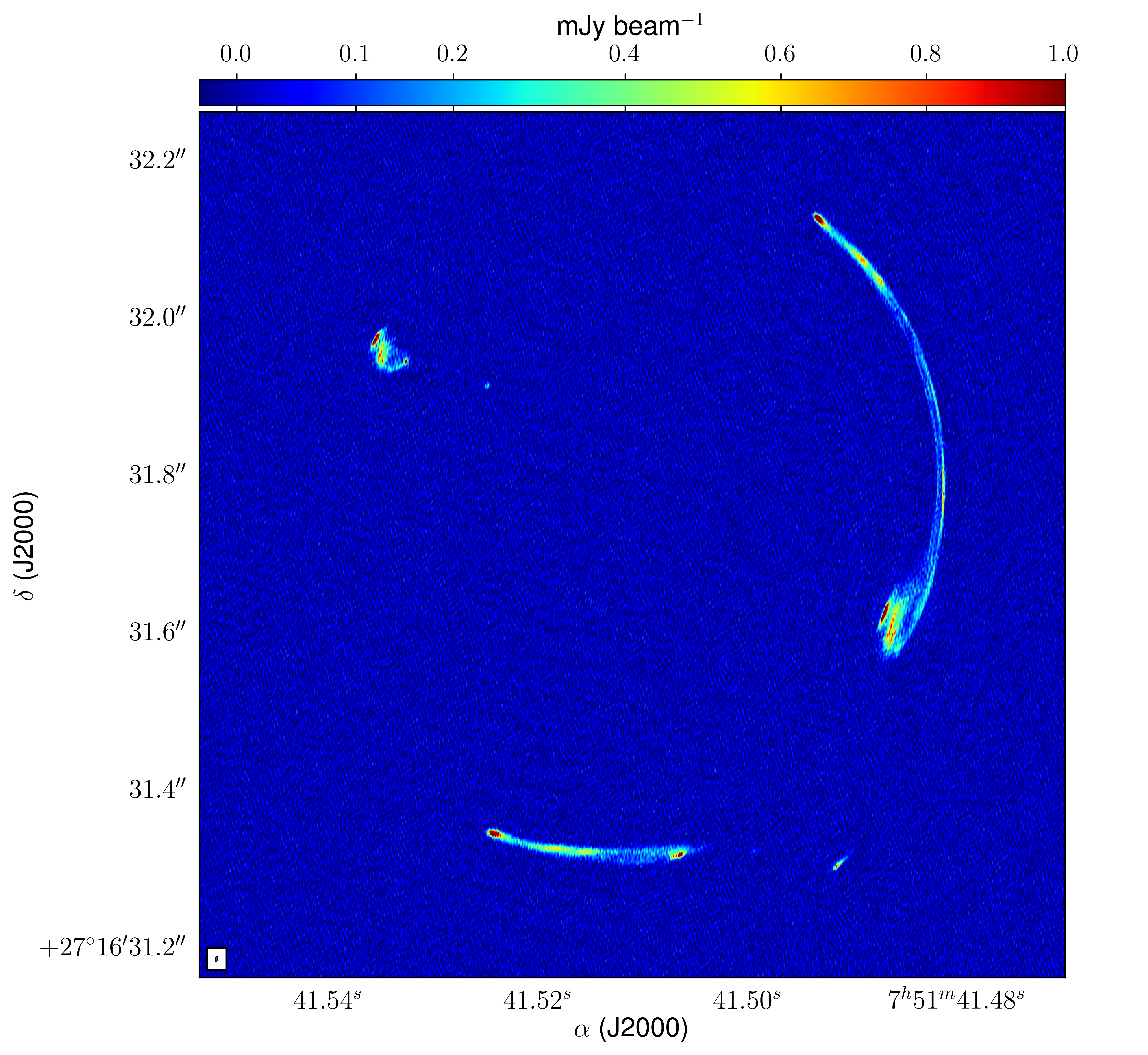} 
\caption{Global VLBI imaging of MG J0751+2716 at 1.65 GHz obtained by using uniform weights and multi-scale {\sc clean}ing in {\sc wsclean}. The off-source rms is $41~\mu$Jy~beam$^{-1}$ and the peak surface brightness is 2.9~mJy beam$^{-1}$. The restored beam is $5.5 \times 1.8$~mas$^2$ at a position angle $-9.8$~deg, and is shown within the white box in the bottom left hand corner.}\label{fig1}
\end{figure*}

\section{Observations}
\label{obs}

In this section, we present the global VLBI observations of MG~J0751+2716 and optical imaging of the field galaxies that we use as additional components to the lensing mass model.

 \begin{table*}
  \centering
  \caption{The measured positions of the various components used for lens modelling, as indicated in Figs.~\ref{fig4} and \ref{fig5}. The observed relative Right Ascension and Declination are determined by performing elliptical Gaussian fits using {\sc jmfit} within {\sc aips}, and are measured with respect to component A$_{1}$ (phased referenced absolute position 07$^h$51$^m$41.487$^s$, +27$^{\circ}$16'31.621"). The position of component D$_7$ is the model-predicted position. The predicted positions from the lens models tested here are also relative to the observed position of component A$_1$. The peak surface brightness (mJy~beam$^{-1}$) is given as a reference, and is not used as a constraint for the lens models. The reported error on the peak surface brightness is the nominal error of the Gaussian fit.}\label{tab1} 
	\begin{tabular}{llllllll}
    	\hline
		\noalign{\vskip 0.15cm}    
		\multirow{2}{*}{ID} & \multicolumn{2}{c}{Observed} & \multicolumn{2}{c}{Model 1} & \multicolumn{2}{c}{Model 2} & I \\
	 	& $\alpha$ (mas) & $\delta$ (mas) & $\alpha$ (mas)  &$\delta$ (mas) & $\alpha$ (mas)  &$\delta$ (mas) &(mJy~beam$^{-1}$) \\
		\cmidrule(lr){1-1}\cmidrule(lr){2-3}\cmidrule(lr){4-5}\cmidrule(lr){6-7} \cmidrule(lr){8-8}
		\noalign{\vskip 0.15cm}    
		A$_1$	& 0.00$\pm$0.01				& 0.00$\pm$0.02	 			& $-$3.27		& $+$2.86		& $-$2.64		& $+$2.56	& 4.91$\pm$0.63 \\
		A$_2$	& $+$6.87$\pm$0.02			& $+$35.60 $\pm$0.20		& $+$2.05  		& $+$27.86		& $+$1.28		& $+$29.81	& 	2.29$\pm$0.67 \\
		A$_3$	& $+$9.99$\pm$0.02			& $+$20.29$\pm$0.10		& $+$12.81		& $+$21.78		& $+$11.34		& $+$24.63	& 2.27$\pm$0.08 \\
		A$_4$	& $+$47.50$\pm0.07$		& $+$21.50 $\pm$1.00	   	& $+$52.89 	& $+$37.59		& $+$64.14		& $+$26.02	& 0.67$\pm$0.02\\ [6pt]	
		B$_1$	& $-$493.31$\pm$0.01		& $-$279.25$\pm$0.03   	& $-$487.28	& $-$277.09	& $-$485.34	& $-$279.18	& 2.78$\pm$0.22	\\
		B$_2$	& $-$414.61$\pm$0.03		& $-$298.87$\pm$0.03 	& $-$410.23  	& $-$292.15	& $-$413.28	& $-$294.04	& 1.20$\pm$0.02	\\
		B$_3$	& $-$379.66$\pm$0.04  	& $-$302.01$\pm$0.20	 	& $-$382.89	& $-$298.69	& $-$387.79	& $-$300.46& 1.20$\pm$0.03	\\
		B$_4$	& $-$261.65$\pm$0.02  	& $-$304.10$\pm$0.04   	& $-$262.34	& $-$306.69	& $-$262.03	& $-$308.44	& 1.96$\pm$0.18	\\
		B$_6$ 	& $-$58.30$\pm$0.30		& $-$318.10$\pm$0.40	 	& $-$58.35		& $-$320.85	& $-$60.34		& $-$321.27	& 0.11$\pm$0.01	\\
		B$_7$ 	& $-$164.37$\pm$0.01 	& $-$300.44$\pm$0.05		& $-$163.19	& $-$299.46	& $-$162.94 	& $-$301.33	& 1.24$\pm$0.06\\	[5pt] 
		C$_1$	& $-$81.45$\pm$0.01		& $+$501.27$\pm$0.02	& $-$83.00		& $+$501.61	& $-$86.97 		& $+$499.99	& 3.25$\pm$0.23	\\
		C$_2$	& $-$29.85$\pm$0.02  		& $+$451.82$\pm$0.13	& $-$25.68  	& $+$456.31	& $-$26.95		& $+$451.90	& 1.53$\pm$0.05 	\\	
		C$_3$	& $-$5.82$\pm$0.03			& $+$424.10$\pm$0.01	& $-$2.31   		& $+$426.19	& $-$2.98 		& $+$421.22 & 1.44$\pm$0.05	\\
		C$_4$	& $+$59.50  $\pm$0.01	& $+$278.50$\pm$0.80	& $+$84.93		& $+$242.38	& $+$85.21		& $+$215.30	& 0.68$\pm$0.05	\\ [5pt]
		D$_1$	& $-$643.05$\pm$0.02 	& $+$346.88$\pm$0.03	& $-$649.74	& $+$346.88	& $-$649.57	& $+$350.76	& 2.48$\pm$0.38	\\
		D$_2$	& $-$639.02$\pm$0.02 	& $+$328.94$\pm$0.05	& $-$643.21	& $+$328.32	& $-$644.16	& $+$326.69	& 1.97$\pm$0.12	\\
		D$_3$	& $-$639.23$\pm$0.02 	& $+$329.81$\pm$0.05	& $-$635.49 	& $+$327.41	& $-$636.86	& $+$326.23	& 	2.04$\pm$0.09 \\
		D$_4$	& $-$607.80$\pm$0.02 	& $+$321.78$\pm$0.02	& $-$606.77    & $+$316.33	& $-$607.30	& $+$315.49	& 1.53$\pm$0.12\\ [5pt]
		D$_6$	& $-$503.82$\pm$0.03 	& $+$290.59$\pm$0.04	& $-$510.54	& $+$291.12	& $-$509.35	& $+$283.97	& 0.52$\pm$0.02	\\
		D$_7$	& $-$574.03		  				& $+$302.87 						& $-$574.03 	& $+$302.87	& $-$574.03	& $+$302.87	& $<$0.01\\
     	\noalign{\vskip 0.15cm}  
     	\hline
	\end{tabular} 
\end{table*} 

\subsection{Very Long Baseline Interferometry data}

MG~J0751+2716 was observed at a central frequency of 1.65 GHz with the global VLBI array on 2012 October 21 for a total time of 18.5~h (project GM070; PI: McKean). The observation comprised 24 antennas from the European VLBI Network (EVN) and the Very Long Baseline Array (VLBA), and included the large ($>50$~m) Lovell, Effelsberg, Robledo and Green Bank telescopes. The scans on the target were about 3~min in duration, which were interleaved by scans of about 2~min on the phase-reference source J0746+273. Several observations of the bright calibrator sources 4C39.25 and DA193 were taken throughout the run for fringe finding during correlation and for the bandpass calibration at the data reduction stage. The data were recorded at 512~Mbits~s$^{-1}$ and correlated at the Joint Institute for VLBI in Europe (JIVE) to produce 8 spectral windows (IFs) with 8 MHz bandwidth and 32 channels each, through both circular polarizations (RR, LL).  A visibility averaging time of 2~s was used. This time and channel resolution limited the effective field-of-view of the observations to about 16 and 10 arcsec, respectively, from the phase centre, which easily encompassed all of the expected structure of the target.

The dataset was initially edited, calibrated and reduced using the EVN pipeline and the Astronomical Image Processing Software ({\sc AIPS}) to produce/apply standard calibration tables. However, during the fringe-fitting process, three antennas (Shanghai, Urumqi and Svetloe) were lost because they had baselines with a signal-to-noise ratio of $<5$. After the initial calibration was completed, new models for the calibrators and target were obtained, which were then used to re-run the fringe-fitting process; this additional step improved the corrections for the residual fringe rates and delays on all antennas, with the exception of Svetloe, which could not be recovered. The phase-referenced dataset for MG~J0751+2716 was then imaged and self-calibrated. A solution interval of 120 to 30~s was used to perform several iterations of phase-only self-calibration. Finally, amplitude self-calibration was applied using at first a long solution interval (lasting the whole observation for each antenna) that was gradually reduced to 30 min to remove any residual uncertainties from the antenna gains. We note that the self-calibration process leads to a global shift in the absolute position of the lensed images by about 1 mas, that is, a fraction of the synthesized beam size. However, this does not affect our gravitational lens modelling, because we use the relative positions of the lensed images, which are not changed by the self-calibration process.

As the system is quite complex and extended, the imaging was performed using multi-scale cleaning,  which is more efficient at modelling extended structures \citep{Cornwell2008,Rich2008}, within the {\sc wsclean} algorithm \citep{Offringa2014}. Our final image of MG~J0751+2716 is presented in Fig.~\ref{fig1}, which was obtained by using uniform weights for the visibilities; the off-source rms is 41 $\mu$Jy~beam$^{-1}$ and the peak surface brightness is 2.9 mJy~beam$^{-1}$.  The restored beam is $5.5 \times 1.8$~mas$^2$ at a position angle $-9.8$~deg east of north. 

\begin{figure}
\centering
\includegraphics[width = \columnwidth]{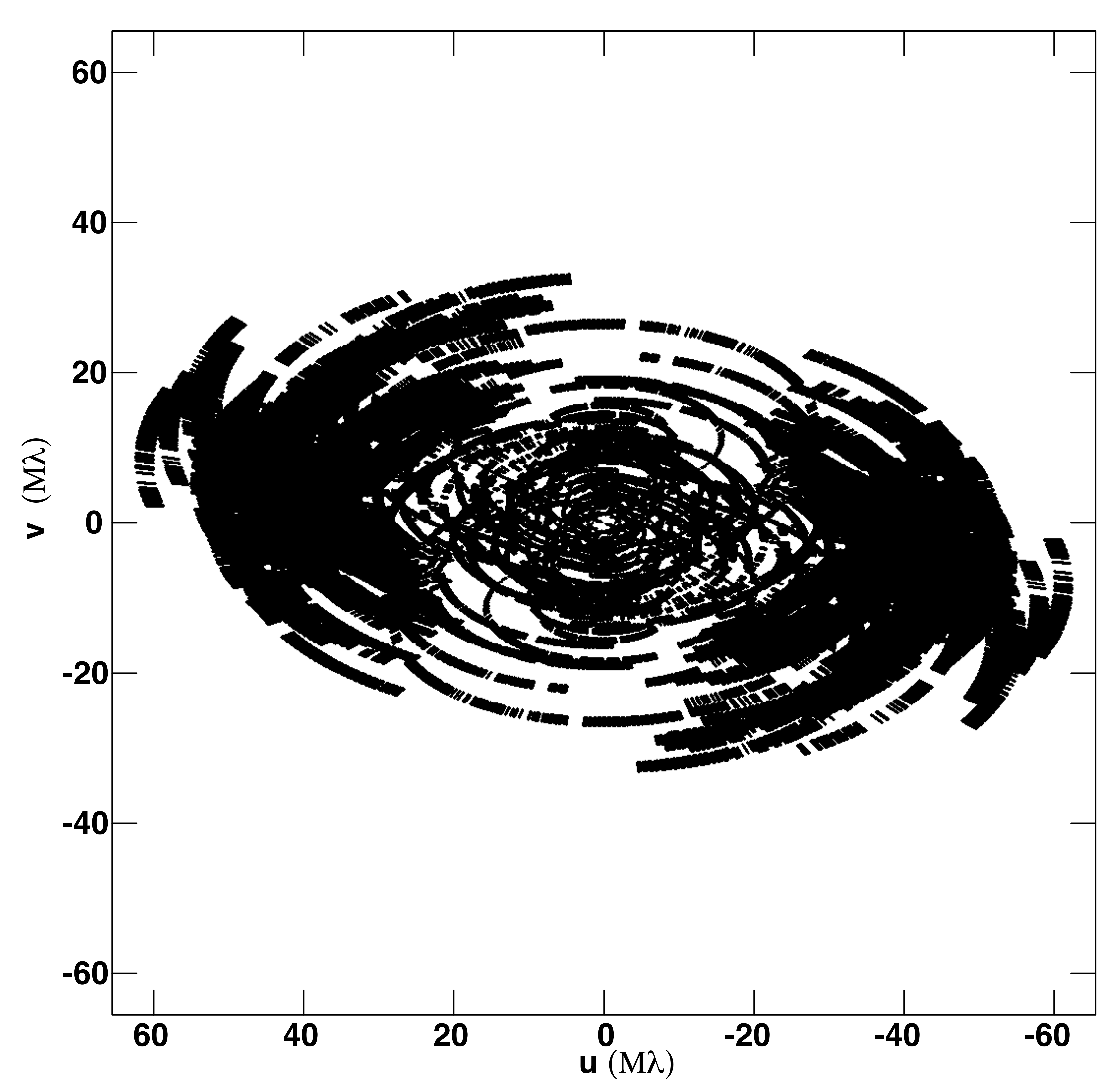}
\caption{The $uv$-coverage of the global VLBI observations of MG~J0751+2716 at 1.65 GHz.}\label{fig2}
\end{figure}

In Fig.~\ref{fig1}, we see that the extended arcs that were previously detected with MERLIN by \citet{Lehar1997} are now resolved into several sub-components that are connected via diffuse jet structure. 
Components A and C are resolved into four sub-components, while components B and D are resolved into 6 sub-components. The pair of merging images (A4 and C4) are highly distorted in the tangential direction with a low flux density emission. Components A(1$\rightarrow$3)  have a similar morphology of components D(1$\rightarrow$3), distorted both in the radial and tangential direction. In addition, the counter image of the doubly-imaged part (components B6 and D6) of the radio source is detected for the first time. We also detect a new source components (B7) at  the 4$\sigma$ level in the doubly-imaged  region. Never before have such extended gravitational arcs been observed at such a high angular resolution. This demonstrates the excellent \textsl{uv}-coverage and surface brightness sensitivity provided by the global VLBI array (Fig. \ref{fig2}), which is fundamental for a detailed study of the structure of extended arcs on mas-scales from objects like MG~J0751+2716. For example, the global VLBI array sensitivity is 2.5 times better than an EVN only observation and 10 times better than a VLBA only observation. The total flux density of MG~J0751+2716 was determined in the image plane by placing an aperture over the area that contains the arcs and the double component, and was found to be $S_{\rm 1.65~GHz} = 350\pm 35$~mJy. 

We conservatively assume an uncertainty on the absolute flux density scale of $\sim10$~per cent, based on the gain and system temperature variations during the observation.

\subsection{\textit{Hubble Space Telescope} data}
\label{sec_HST}

\begin{figure*}
\centering
\includegraphics[scale = 0.027]{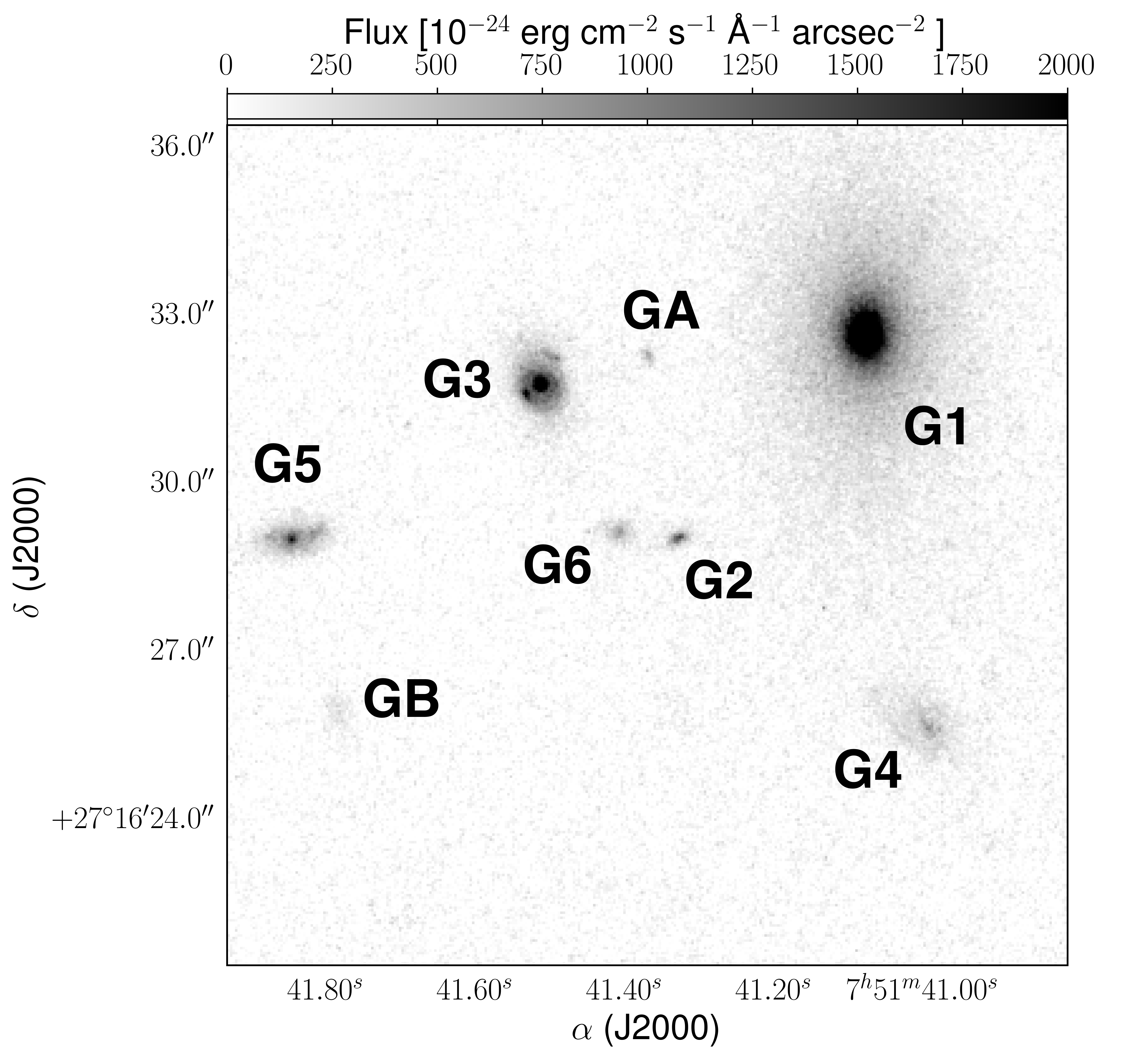}
\includegraphics[scale = 0.37]{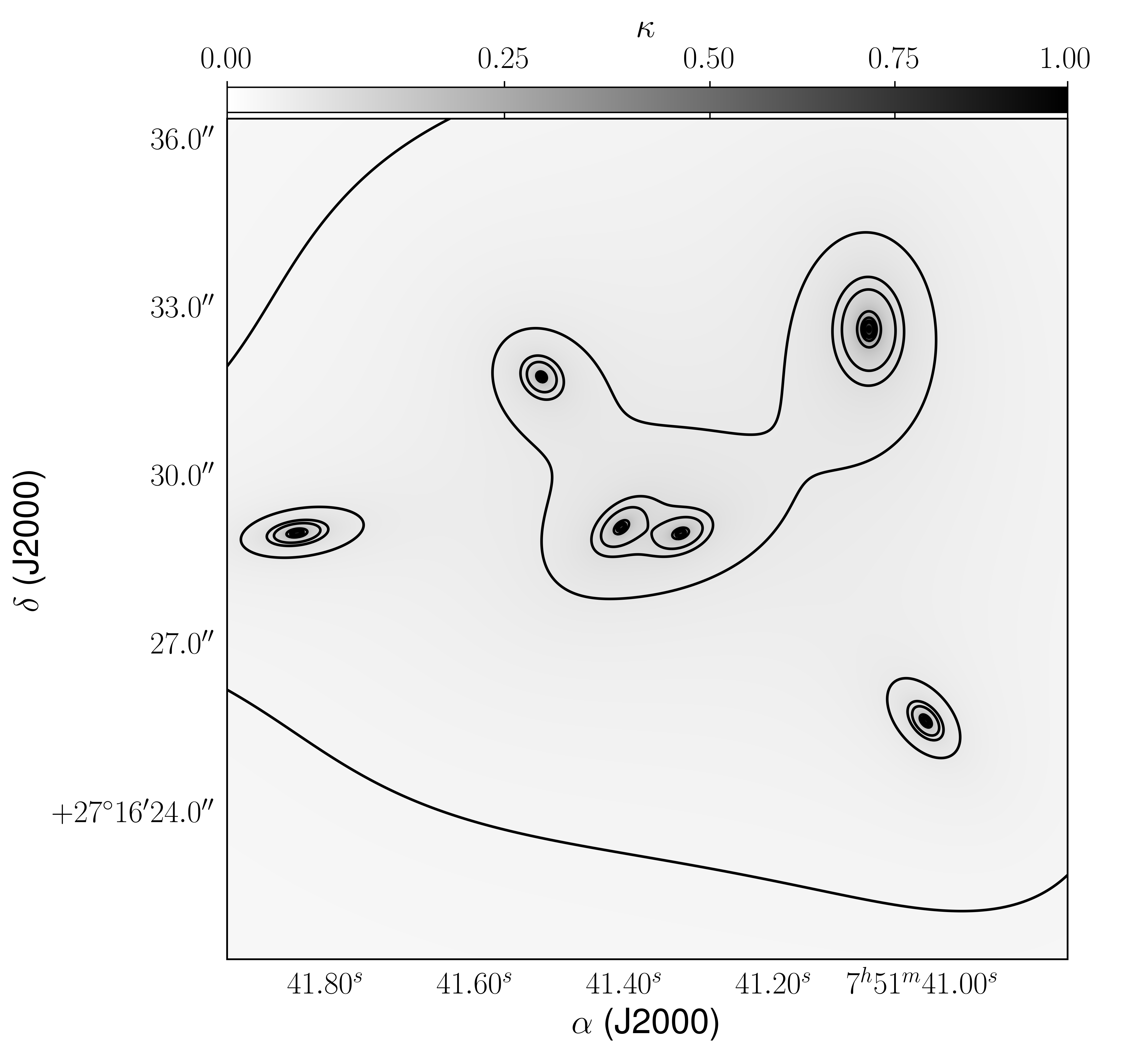}
\caption{(Left) {\it HST} WFPC2 F814W image of the MG J0751+2716 group of galaxies, where G3 is the main lensing galaxy. The field of view of this image is  $15\times15$~arcsec$^2$. The nomenclature for the galaxies follows \citet{Lehar1997} and \citet{Alloin2007} for G1 to G6, which are all spectroscopically confirmed group members \citep{Momcheva2006}.  (Right) The dimensionless convergence map for Model 2 (see Section~\ref{model}), showing the combined contribution of the individual group galaxies to the mass model. GA and GB are not included in Model 2 since they do not have spectroscopic information. The contours show regions of iso-convergence for $\kappa =  0.05$, 0.1, 0.2 and 0.4.}
\label{fig3}
\end{figure*}

As MG~J0751+2716 is known to be gravitationally lensed by a group of galaxies, we use high resolution optical imaging from the {\it HST} to provide the relative positions, ellipticities and position angles of the group galaxies as an input to our mass modelling, The archival optical observations of MG J0751+2716 were obtained as part of the CASTLES program (GO-7495; PI: Falco) using the Wide Field Planetary Camera 2 (WFPC2) through the F814W filter. These  observations were processed in {\tt astrodrizzle} within the {\sc iraf} package using standard procedures. The final drizzled image has a pixel scale of 0.045 arcsec~pixel$^{-1}$ and is shown in Fig. \ref{fig3}. 
In order to estimate the relative position and magnitude of the galaxy group members in this {\it HST} image, we use the software {\sc sextractor} \citep{Bertin1996}. 

\section{Lens Modelling}
\label{model}

\begin{figure*}
\centering
\includegraphics[scale = 0.75]{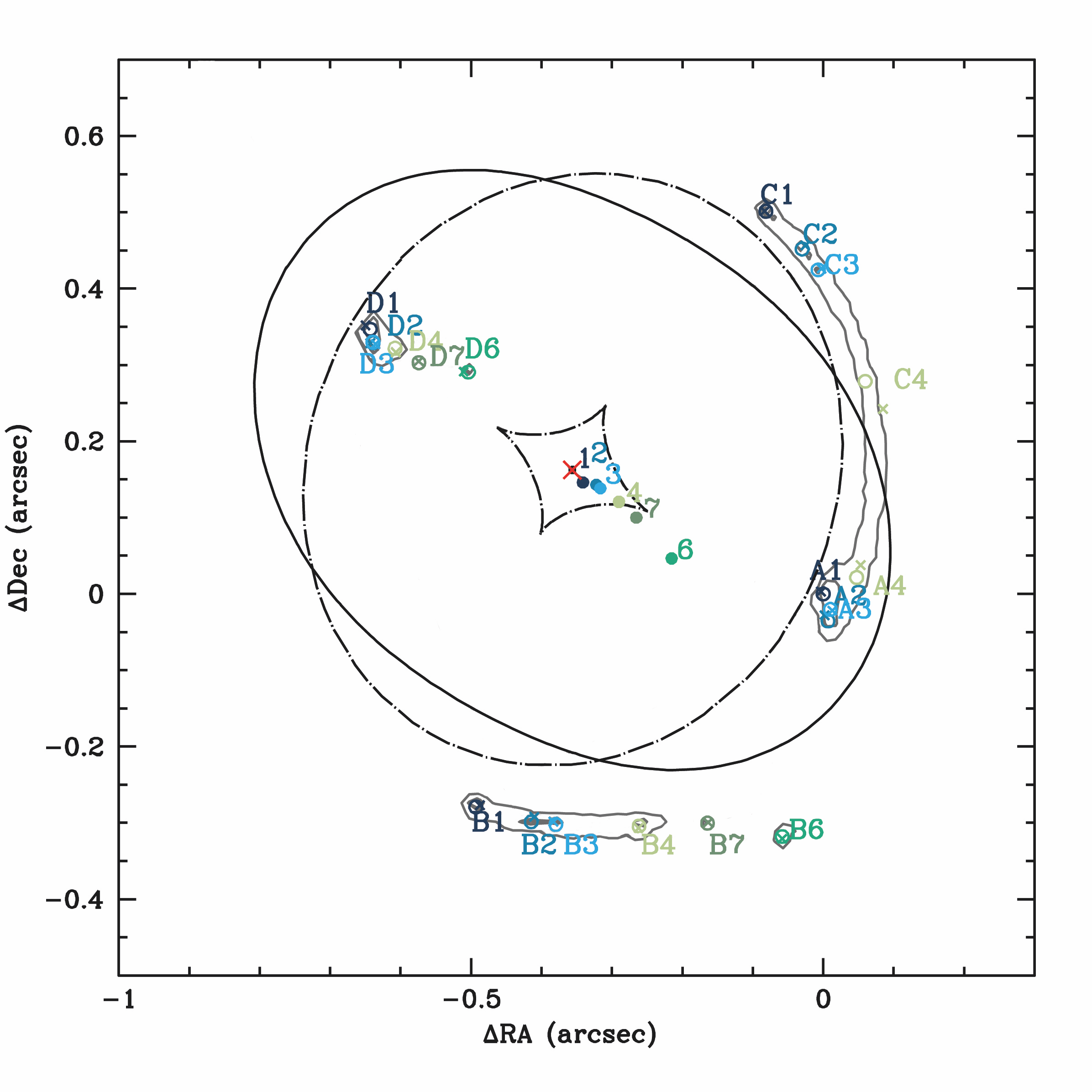}
\caption{Model 1 for MG J0751+2716. The model consists of an ellipsoid power-law mass distribution for the main lensing galaxy G3 (red cross), with an external shear. The observed positions are the open circles and the predicted positions are represented by the crosses, with all positions given relative to component A1. Each colour corresponds to a different background source component (filled circles). The lens critical curve is shown by the thick line; the dashed line shows the source plane caustics. The grey lines are the $3\sigma$ contours of the extended emission detected from our global VLBI imaging, for reference. The mass model parameters and their uncertainties are given in Table \ref{tab_lensparameters}.}
\label{fig4}
\end{figure*}
  
\begin{figure*}
\centering
\includegraphics[scale = 0.75]{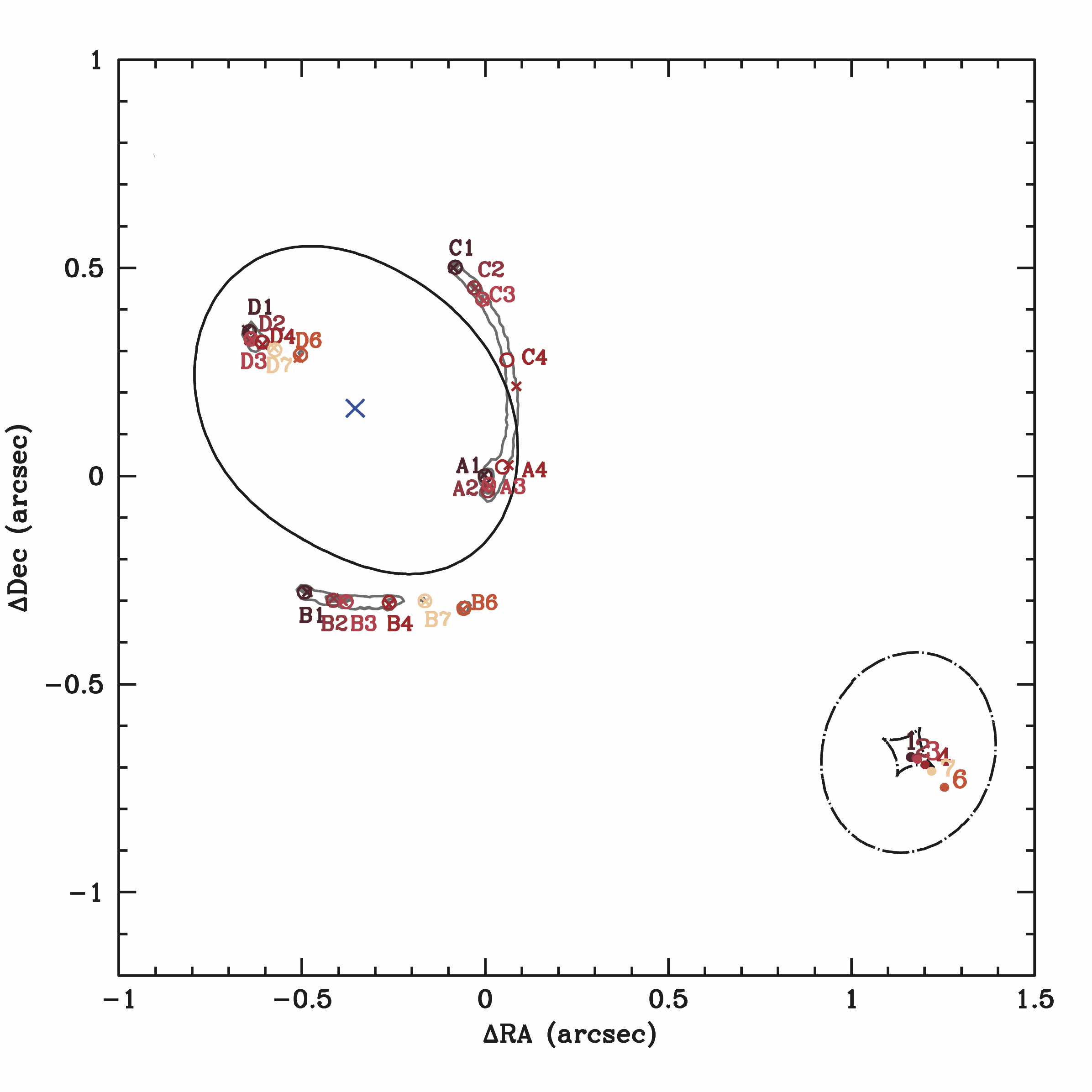}
\caption{Model 2 for MG J0751+2716. The model consists of an ellipsoid power-law mass distribution for the main lensing galaxy G3 (blue cross) and five SIEs for the group galaxies, with an external shear. The observed positions are the open circles and the predicted positions are represented by the crosses, with all positions given relative to component A1. Each colour corresponds to a different background source component (filled circles). The lens critical curve is shown by the thick line; the dashed line shows the source plane caustics. The grey lines are the $3\sigma$ contours of the extended emission detected from our global VLBI imaging, for reference. The mass model parameters and their uncertainties are given in Table \ref{tab_lensparameters}.}
\label{fig5}
\end{figure*}

The high angular resolution of the data and the wealth of extended structure that has been detected from MG~J0751+2716 can provide many  constraints to the gravitational lens mass model. In principle, such an analysis should include all of the structure that is observed as a constraint. However, this requires producing a model for the lens and background source that fits the observed data, which in this case are the visibilities. 
Since producing such a model is computationally expensive, we instead start with a simpler case: we generate a model by matching the conjugate positions of compact sub-components seen in each multiply-imaged arc.
The more sophisticated grid-based modelling of the data will be presented in a future paper.

\subsection{Parametric lens modelling}

We use the publicly available code {\sc gravlens} \citep{Keeton2001a,Keeton2001} to model the compact radio components. As primary constraints to the starting model, we use the relative positions of four source components that are quadruply imaged (from the four extended arcs) and two components that are doubly-imaged. The relative positions of these components are listed in Table \ref{tab1} and are shown in Figs.~\ref{fig4} and \ref{fig5}.  We also give the peak surface brightness of each component in Table \ref{tab1}, but these are not used as model constraints for all of the components except of the two doubly-imaged components; since they are both compact and fairly isolated, we use their relative flux-ratios as additional model constraints.  

We first identify groups of lensed images that correspond to the same background source component. Next, we fit each of these image components with a single elliptical Gaussian model in the image plane using the task {\tt jmfit} within {\sc AIPS} to determine their position. In this way we take into account the extended nature of each source component and obtain a weighted position for the individual lensed images. We apply a different approach for the groups of image components 4 and 7, which are either highly distorted or not detected in more than one of the lensed images. We instead use these image components as a test for our best model, as opposed to using them as constraints. For source component 4 we could confidently identify the more compact image components B4 and D4, while A4 and C4 are highly distorted by the lens (e.g., see Fig. \ref{fig1}). Therefore, we use the model-predicted position for A4 and C4 to fit a Gaussian in the image plane and obtain the weighted average position for these two images, which we show in Figs.~\ref{fig4} and \ref{fig5}. We follow the same method for finding the predicted position and flux density of image component D7. The relative model-predicted flux density of image component D7 is about 0.16 of image component B7, which is detected at the 4$\sigma$ level. The flux density of image component B7 is $223\pm55~\mu$Jy. Therefore, the model-predicted flux density for D7 is $\sim 36~\mu$Jy, which would correspond to a non-detection in our image, and is consistent with the data.

The positional uncertainty on each component is calculated in the standard way by using the major and minor axes of the elliptical Gaussian determined with {\tt jmfit}, and their signal-to-noise ratio (calculated using the peak surface brightness given in Table \ref{tab1} and the rms of the uniform weighted image). We assume larger positional uncertainties for the relative declination of the components A4 and C4, because they show a significant distortion in the declination direction. With this assumption, they do not significantly effect the final $\chi^2$ of the lens model.

Moreover, since we use the relative position of the images as constraints to the lens modelling, we can neglect the systematic errors in the estimate of the absolute positional accuracy.  The main lensing galaxy (G3 following the nomenclature of \citealt{Lehar1997}; see also Fig. \ref{fig3}) is known to be an elliptical galaxy, as shown from the surface brightness profile at near-infrared and optical wavelengths (e.g. \citealt{Lehar1997, Spingola2018}) and from the optical spectrum \citep{Tonry1998}. Therefore, we model the lensing potential as a power-law ellipsoid density profile, which has been a remarkably good fit to early-type lensing galaxies that were observed with the {\it HST} and Keck adaptive-optics imaging \citep{Lagattuta2012,Vegetti2014,Oldham2017}. However, in the case of MG~J0751+2716, there is known to be a significant perturbation to the lensing potential, which could be due to the group of galaxies associated with the main lensing galaxy \citep{Lehar1997, Momcheva2006, Alloin2007}. Therefore, we take this into account by considering two models; Model 1 includes the main lensing galaxy and an additional external shear component, while Model 2 includes the lensing galaxy, external shear, and five additional haloes representing all spectroscopically confirmed group-member galaxies (see Fig. \ref{fig3} and Sect. \ref{sect_results}).

As there is a known degeneracy between the ellipticity and the external shear, we perform the optimization by using the following method. We first find the optimal position angles and  lens strength for G3 by using the values obtained by \citet{Lehar1997} as an initial guess. We then keep the ellipticity and shear fixed, and optimize for the position angle of each. For the next step, we optimize for all of the parameters by keeping the ellipticity and shear fixed. Finally, we leave all parameters free and optimize for everything. From the positions of the four quadruply-imaged components and the two doubly-imaged components measured from our global VLBI observations, we obtain 36 positional constraints that are also probing the lensing potential over a large region (e.g., see Fig.~\ref{fig4}).  

For Model 1, we have 12 variables to describe the 6 source positions, and the lens mass model has 8 variables [mass scale ($b$); position ($x_L, y_L$); ellipticity ($e$), position angle ($\theta$); power-law slope ($\gamma$), external shear ($\Gamma$) and its position angle ($\Gamma_\theta$)]. For Model 2, we have the same number of model parameters, except that we fix the position of the galaxies relative to G3 based on their centroid positions from {\it HST} imaging (G3 position is free; $x_L, y_L$), and we fix the mass of the galaxies relative to G3 (which is free; $b$) using their relative optical magnitudes and scaling relations according to their Hubble type (Faber--Jackson or Tully--Fisher, e.g. \citealt{McKean2005,More2008}). We infer the best fit parameters and the uncertainty on the values (68 per cent confidence level) from a Markov-Chain Monte Carlo (MCMC) sampler implemented in {\sc gravlens}.

\subsection{Results}
\label{sect_results}

The results for Model 1 and 2 are given in Table~\ref{tab_lensparameters}, and the observed and model-predicted image positions, with the critical and caustic curves, are shown in Figs.~\ref{fig4} and \ref{fig5}, respectively. The residual image positions between the data and models are shown in Fig. \ref{fig6}. The probability density distribution for each parameter is shown in Figs.~\ref{fig7} and \ref{fig8} for Model 1 and 2, respectively. We now briefly describe the results for the two models considered here.

\subsubsection{Model 1 -- A single lensing galaxy and external shear}

We first test whether the image configuration can be explained with a simple ellipsoidal power-law density profile, with an external shear. This model is similar to previous single-lens models that assumed an isothermal density profile \citep{Lehar1997}, but has much tighter constraints on all of the parameters given the increased precision provided by the mas-scale resolution of our global VLBI observations.

The Einstein radius of G3 is found to be $b = 0.4025^{+0.0007}_{-0.0008}$ arcsec, with an  ellipticity of $e = 0.159\pm 0.001$ at a position angle of $36\pm3$~deg (east of north). The ellipticity and position angle of the surface brightness profile as measured in the {\it HST} F814W imaging using the software {\sc galfit} (\citealt{Peng2010}) are $0.35 \pm 0.04$ and $16\pm 2$~deg, respectively. This misalignment of $\sim$20 deg between the mass distribution and the light profile is not unexpected due to the significant external perturbation from the other galaxies in the group, which can affect the shape of the lensing potential \citep{Keeton2000,Kochanek2001}. This level of misalignment has also been found in other lensing groups (e.g. CLASS~B2108+213; \citealt{McKean2010,More2009}) and in lensing systems with a substantial external shear (e.g. \citealt{Gavazzi2012}). Indeed, there is a significant amount of external shear needed by this model ($\Gamma = 0.084\pm0.005$), and its position angle ($\Gamma_{\theta} = 79\pm2$~deg) suggests that the BCG of the group, to the west of the main lensing galaxy, is the principal cause of this external perturbation (see Fig.~\ref{fig3}). Such a high external shear is typical for galaxies lying in a group or cluster of galaxies (e.g. \citealt{Keeton1997,Oguri2005,Auger2007}). Finally, the best-fit model suggests a power-law density slope for the main lensing galaxy that is steeper than isothermal ($\gamma =2.08\pm0.02$) at the $4.2\sigma$ level. 

This model does not provide a satisfactory fit to the observed images, as shown in Table \ref{tab1} and Fig.~\ref{fig6}. The most difficult images to fit are A4 and C4, because they are highly distorted and, therefore, are not properly represented by a single elliptical Gaussian component; any small change in the position of the source-component 4 will have a significant change in the position of image-components A4 and C4 due to its position relative to the caustic. Also, these images have larger uncertainties on their positions with respect to the other images (Table \ref{tab1}), therefore they are not adequately constrained during the optimization and they do not dominate the $\chi^2$ minimization.  Moreover, we also find for the other images a significant mismatch between their observed and predicted positions, with respect to the astrometric uncertainty, with mean offsets of the order of $\sim200\sigma$.


\begin{table*}
 \centering
 \caption{\small The minimum-$\chi^2$ parameters of the two parametric lens models for MG J0751+2716 presented in this paper: $b$ is the lens strength (arcsec), $x_L$ and $y_L$ are the positions in Right Ascension and Declination relative to A1 (arcsec), $e$ is the ellipticity, $\theta$ is the position angle of the ellipticity (east of north, degrees), $\Gamma$ is the external shear strength and $\Gamma_{\theta}$ is the external shear position angle (east of north, degrees). The density slope of the ellipsoidal power-law mass distribution is given by $\gamma$, where for an isothermal profile $\gamma = 2$. For Model 2, the positions and ellipticities are fixed to the optical parameters for the group galaxies and their $\gamma$ is fixed to the isothermal case. Also, for Model 2, the Einstein radius of the group galaxies relative to G3 is fixed based on their relative optical fluxes and using the Faber--Jackson or Tully--Fisher relation, depending on Hubble type: G1 and G3 are early-type galaxies, the others are late-type galaxies. For G3 we report the best set of parameters recovered via the minimization with {\sc gravlens} (Best) and the average values with relative 95 per cent limits assessed by the MCMC chains implemented in {\sc gravlens} (Mean).} \label{tab_lensparameters}
\begin{tabular}{llllllllllll}
	\hline
	 \noalign{\vskip 0.15cm}
	 Par. &  \multicolumn{3}{c}{Model 1} & \multicolumn{8}{c}{Model 2} \\
	 \noalign{\vskip 0.15cm}
	\cmidrule(lr){1-1} \cmidrule(lr){2-4}\cmidrule(lr){5-12}
	 &   \multicolumn{3}{c}{G3} &  \multicolumn{3}{c}{G3} & G1 & G2 & G4 & G5 & G6 \\
	 \cmidrule(lr){2-4}\cmidrule(lr){5-7} \cmidrule(lr){8-12}
 	& Mean & $\sigma_{95 \%}^{\rm mean}$ & Best & Mean &$\sigma_{95 \%}^{\rm mean}$ &  Best & & & & &   \\
 	 \noalign{\vskip 0.2cm}
  	$b$ &  $0.40249$ & $^{+0.00074}_{-0.00081}$ & $0.39810$ & $0.3073$ & $^{+0.0021}_{-0.0022}$ &$0.3136$ & $\equiv0.720^{+0.098}_{-0.096} $ & $\equiv0.250^{+0.110}_{-0.120}$ &  $\equiv0.320^{+0.170}_{-0.160}$& $\equiv0.061^{+0.041}_{-0.041}$ &   $\equiv0.265^{+0.052}_{-0.050}$\\ [3pt]
  	$x_L$ 	& $-0.3530$ & $^{+0.0011}_{-0.0011}$ & $-0.3561$ & $-0.35482$ & $^{+0.00018}_{-0.00018}$ & $-0.35448$  & $\equiv+5.447$	& $\equiv+2.120$ & $\equiv+6.530$ & $\equiv-4.807$ & $\equiv+1.066$ \\ [3pt]
  	$y_L$ & $0.1594$ & $^{+0.0010}_{-0.0010}$ & $0.1624$ & $0.16240$ & $^{+0.00019}_{-0.00019}$ & $0.16273$ & $\equiv+1.044$ & $\equiv-2.594$ & $\equiv-6.014$& $\equiv-2.642$ &  $\equiv-2.472$	\\[3pt]
  	$e$ & $0.159$ & $^{+0.014}_{-0.013}$ & $0.2269$ &  $0.1605$ & $^{+0.0080}_{-0.0085}$ &$0.1896$	& $\equiv0.34$ & $\equiv0.40$ &  $\equiv0.40$	 &  $\equiv0.60$	& $\equiv0.40$ \\[3pt]
  $\theta$  & $35.7$ & $^{+3.4}_{-3.3}$ & $49.0$ & $38.9$ & $^{+1.9}_{-2.0}$ & $44.9$ & $\equiv+1.0$ &  $\equiv-70.0$& $\equiv+40.0$	&   $\equiv-82.0$ &  $\equiv-50.0$\\[3pt]
  	$\Gamma$  & $0.0837$ &  $^{+0.0049}_{-0.0053}$ & $0.06605$ & $0.0343$ & $^{+0.0026}_{-0.0026}$ & $0.03109$	& & & & & \\[3pt]
  	$\Gamma_{\theta}$ & $79.2$ & $^{+1.6}_{-1.5}$  & $78.99$ &  $-61.8$ & $^{+3.4}_{-3.7}$ 	&$ -59.8$							& & & & & \\[3pt]
 	$\gamma$ &$2.079$ &$^{+0.019}_{-0.019}$ & $2.008$ &  $2.157$ & $^{+0.023}_{-0.023}$ &$2.078$ 				& $\equiv2.0$& $\equiv2.0$ &  $\equiv2.0$ & $\equiv2.0$& $\equiv2.0$   \\[3pt]
 	\noalign{\vskip 0.1cm}
	 \cmidrule(lr){2-4}\cmidrule(lr){5-12}
 	 \noalign{\vskip 0.15cm}
	$\chi^2_{\rm red}$ & \multicolumn{2}{c}{1.9} & \multicolumn{7}{c}{1.4} \\
 	\noalign{\vskip 0.15cm}
	\hline 
   \end{tabular}  
\end{table*}

\subsubsection{Model 2 -- A lensing group of galaxies and external shear}

For our second model, we explore a more realistic lens mass model for MG J0751+2716 in which we take into account the individual members of the group of galaxies explicitly. The mass distribution of the entire group is parametrized using an ellipsoidal power-law density profile for G3 (the main lensing galaxy), with five SIEs to represent each member of the group of galaxies (G1, G2, G4, G5 and G6), plus an external shear component.
Their position and mass scales relative to G3 are fixed, as also is their ellipticity and position angle, based on the {\it HST} imaging (Table \ref{tab_lensparameters}). A convergence map for this model is shown in Fig. \ref{fig3}, while the marginalized posterior probability distribution function for the lens model parameters is shown in Fig.~\ref{fig8}.

The immediate difference between Model 1 and 2, is a lowering of the Einstein radius of G3 to $b = 0.307 \pm 0.002$~arcsec and a lowering of the external shear to $\Gamma = 0.034\pm 0.003$, as expected since we are now accounting for the external convergence of the system due to the group of galaxies.  
The change of the shear position angle to $\Gamma_{\theta} = -61.8 \pm 3.5$~deg can be attributed to the added complexity of the galaxy environment. The ellipticity of G3 is found to increase to $e = 0.1605\pm0.0002$, and the misalignment between the optical surface brightness profile and the gravitational lensing mass profile is still of the order of $20$~deg. It is not clear whether this misalignment is due to some additional mass structure that is not included in the mass model (see Section~\ref{disc} for discussion), or whether this is evidence for some interaction in the group environment that has affected the G3 dark matter halo to a larger extent than the stellar component. Finally, the power-law of the ellipsoidal density profile for G3 has become even more super-isothermal, with $\gamma = 2.16 \pm 0.02$, at the $6.8\sigma$ level.

\begin{figure*}
\centering
\includegraphics[scale= 0.02]{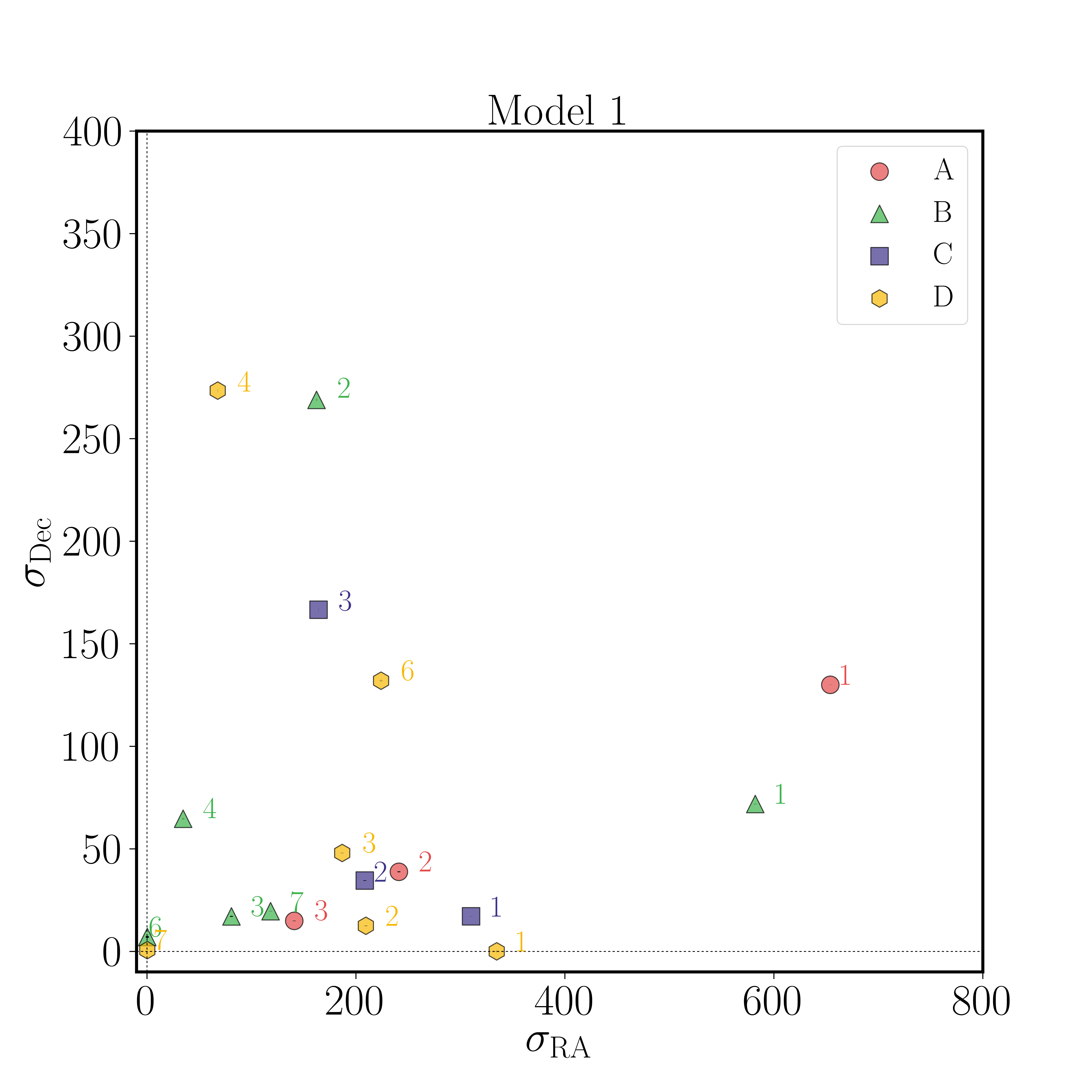}
\includegraphics[scale= 0.02]{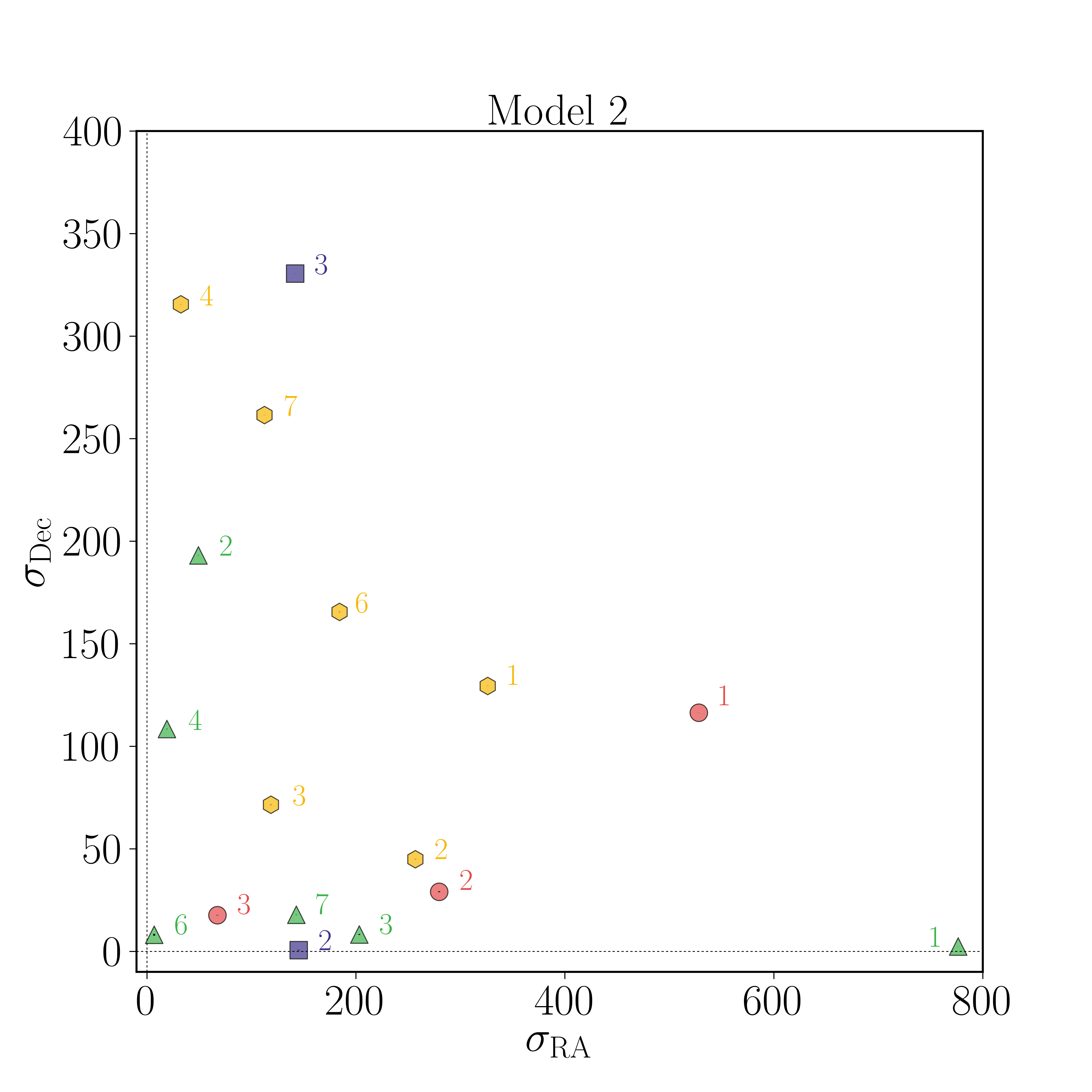}\\
\includegraphics[scale= 0.02]{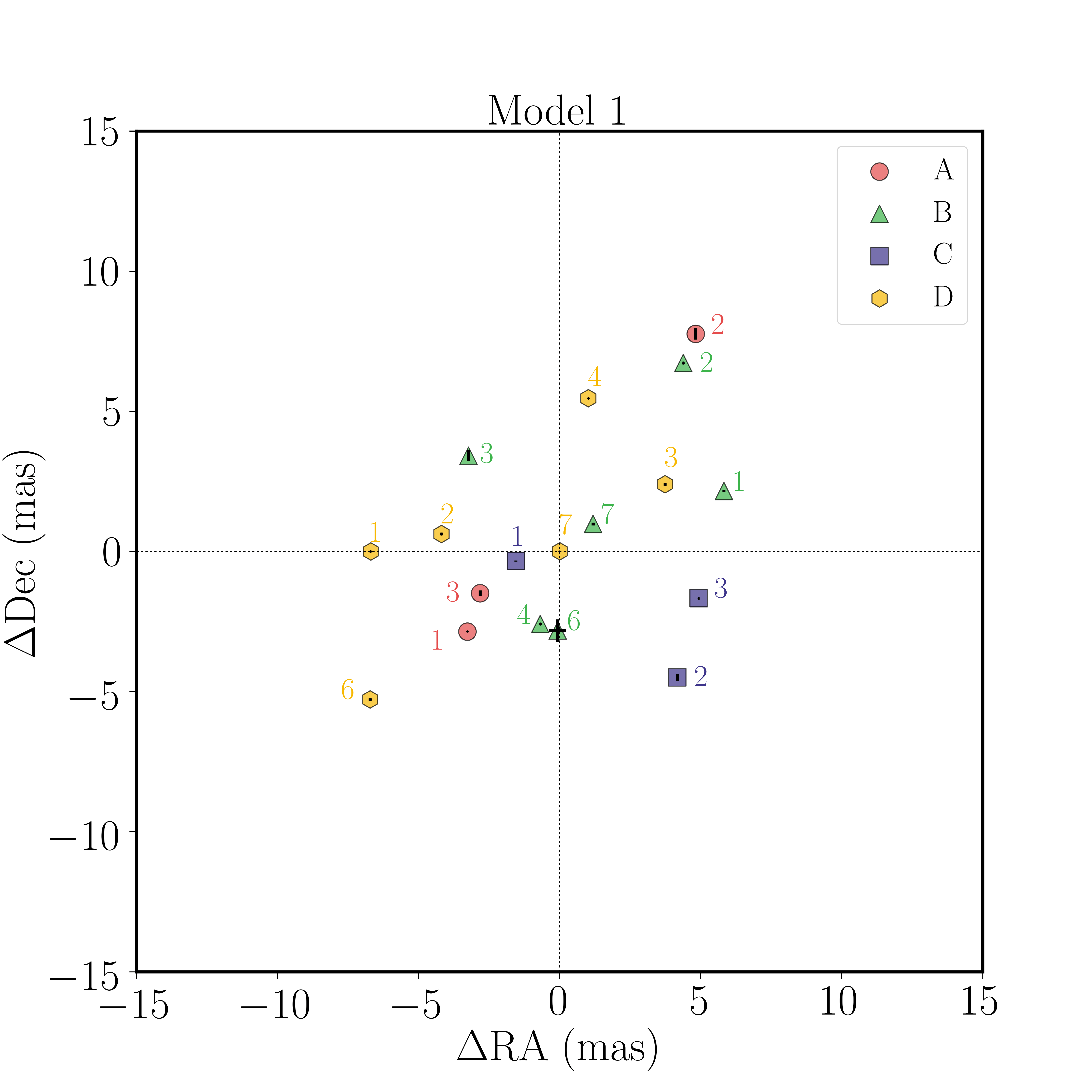}
\includegraphics[scale= 0.02]{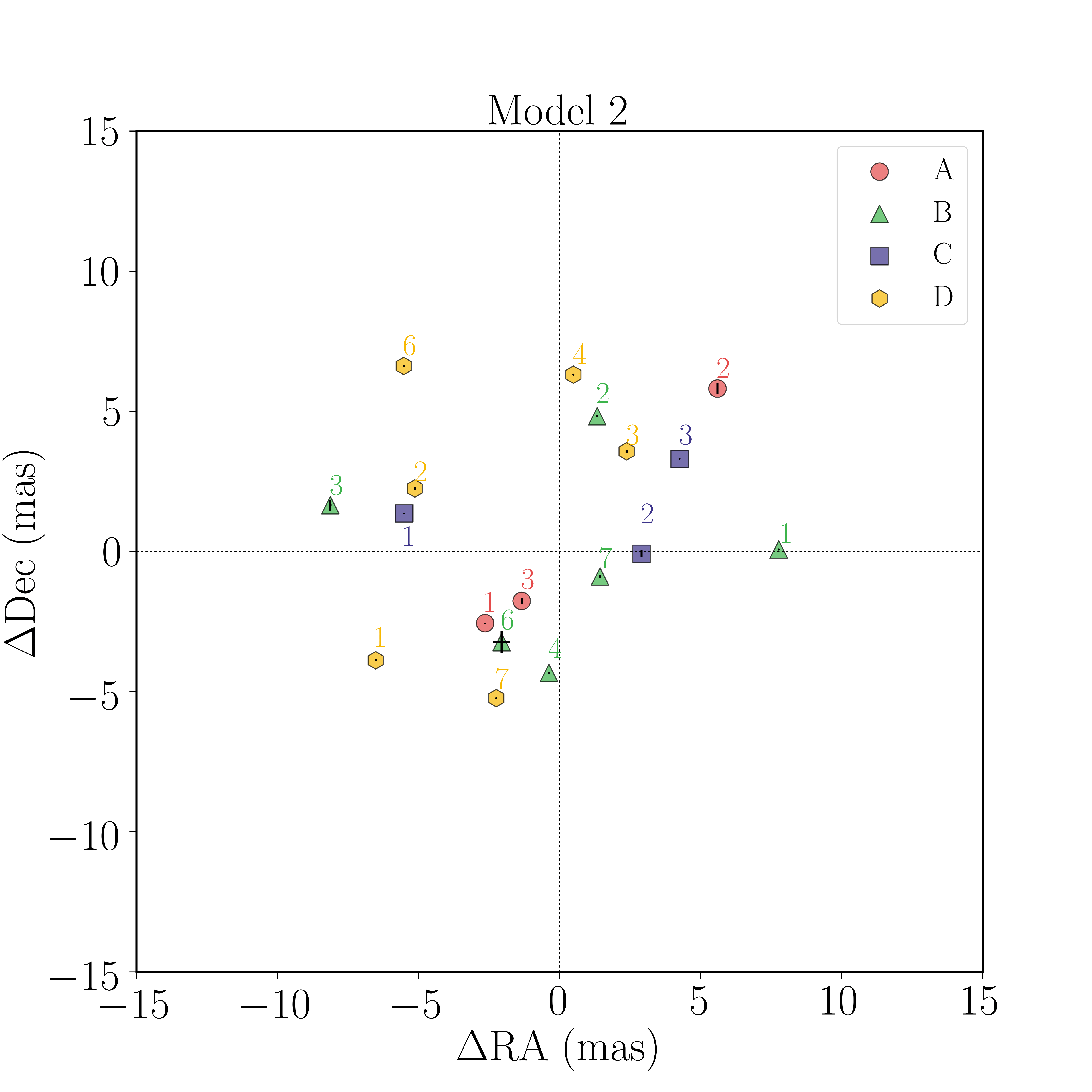}
\caption{The offset between the observed and the model-predicted positions in units of sigma (upper) and mas (lower) for Model 1 (left) and Model 2 (right). Each colour and symbol represents a different group of lensed images as indicated in the legend on the top-right of each panel. The error bars are shown in black and the two black dashed lines indicate the no offset position.} \label{fig6}
\end{figure*}

\begin{figure*}
 \centering
	\includegraphics[scale = 0.094]{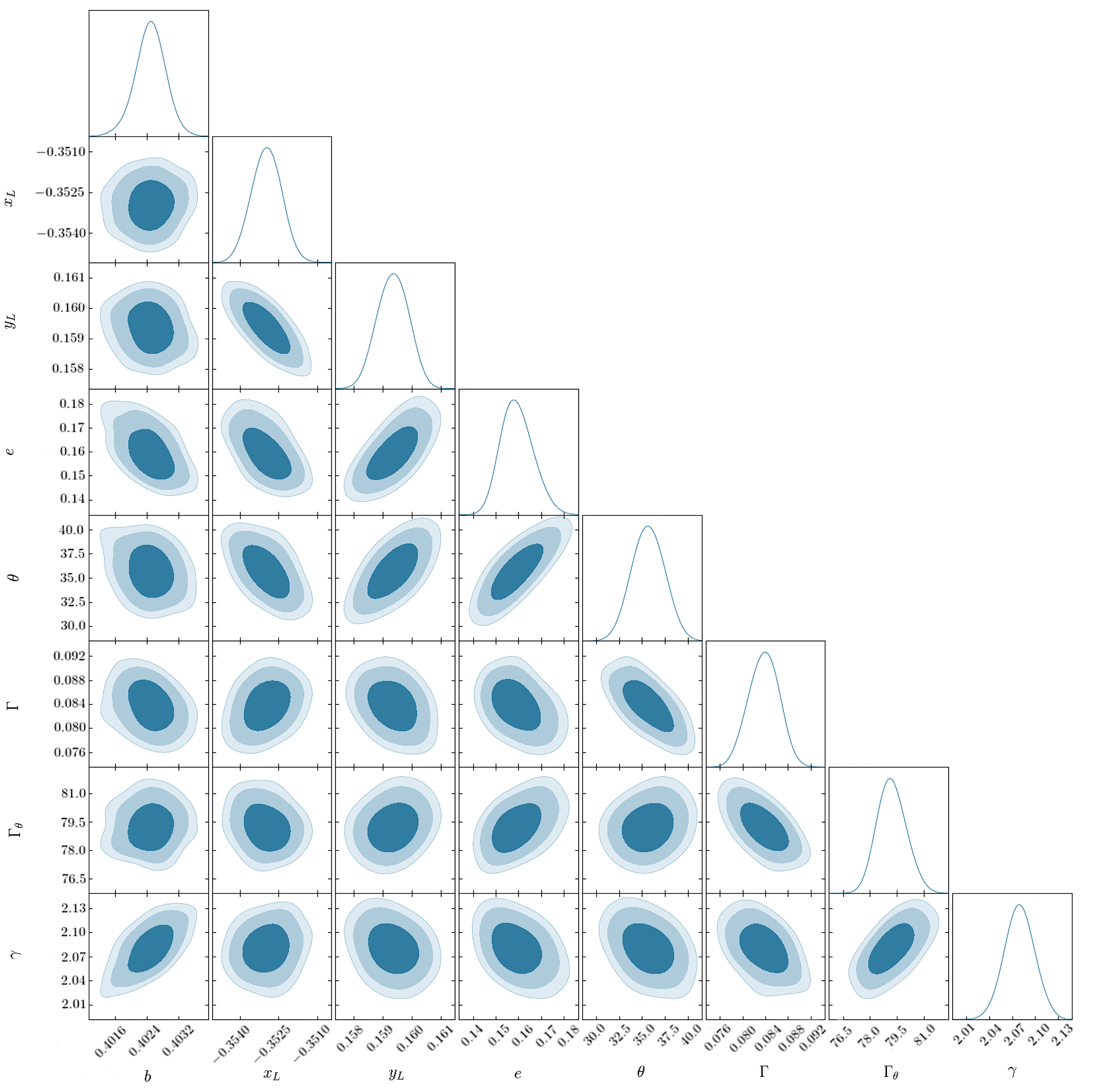}
   \caption{The marginalized posterior probability distribution function (PDF) for the lens model parameters of Model 1. The contours enclosing the 99, 95 and 68 percentiles indicate the distribution between two parameters of the lens model. The PDF of each   parameter is shown at the top of each column. The meaning of the parameters, the maximum-likelihood model values for each parameter, and their uncertainties, are presented in Table \ref{tab_lensparameters}.}\label{fig7}
\end{figure*}

\begin{figure*}
 \centering
	\includegraphics[scale = 0.43]{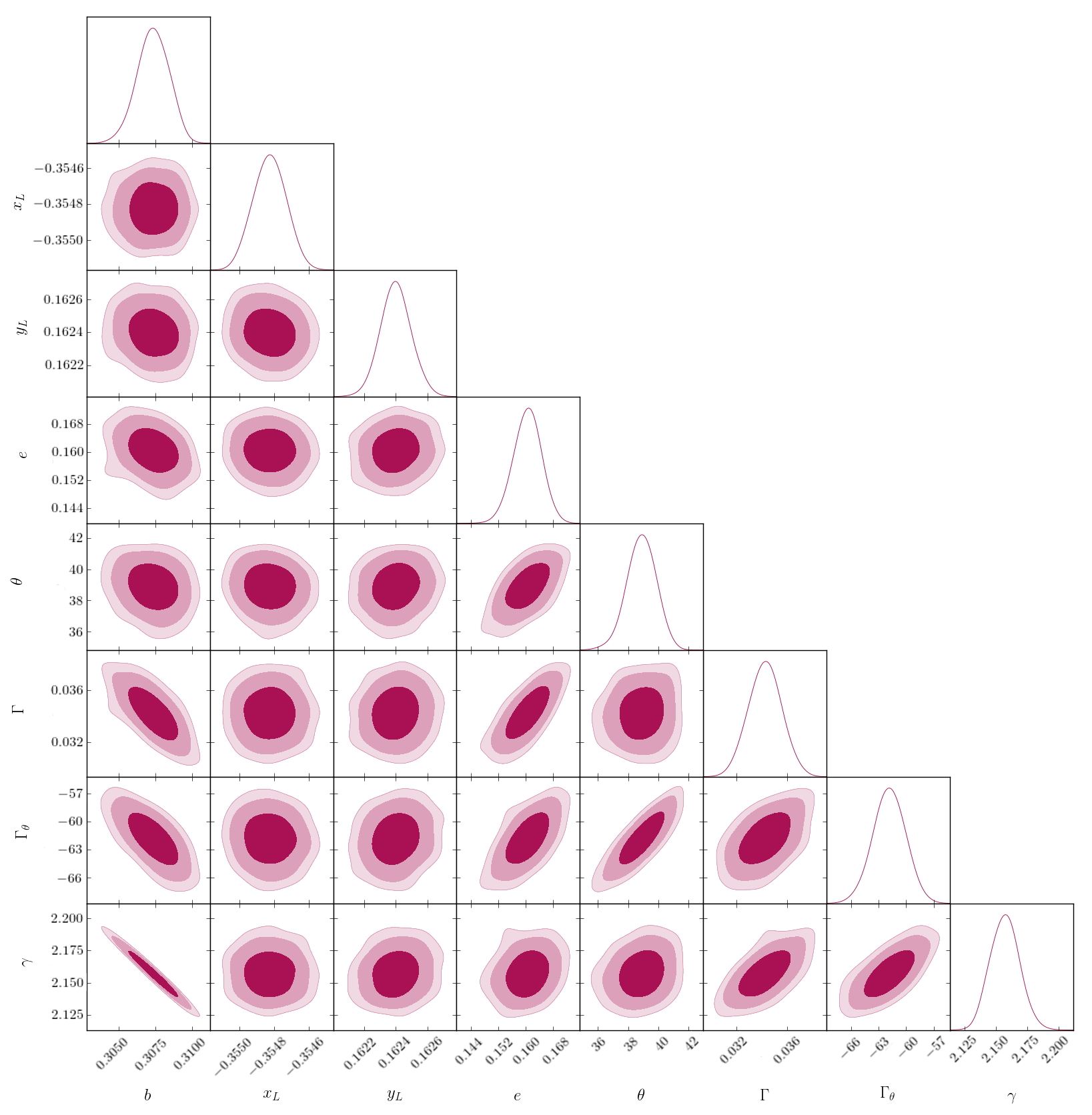}
   \caption{The marginalized posterior probability distribution function (PDF) for the lens model parameters of Model 2. The contours enclosing the 99, 95 and 68 percentiles indicate the distribution between two parameters of the lens model. The PDF of each   parameter is shown at the top of each column. The meaning of the parameters, the maximum-likelihood model values for each parameter, and their uncertainties, are presented in Table \ref{tab_lensparameters}.}\label{fig8}
\end{figure*}

Similar to Model 1, we find that there are significant deviations between the observed and model-predicted positions of the image-components (1--7 mas, 
that correspond to $\sim 200 \sigma$ offsets on average; see Table \ref{tab1} and Fig. \ref{fig6}) and the positional offsets of Model 2 are 1 per cent larger than those of Model 1.  We have carried out some additional tests of Model 2 to investigate what other mass structures could account for these differences, although in general, we find that the data are not sufficient to constrain these additional mass components, so they are not formally part of Model 2.

We first attempted to constrain a possible common halo for the group. Such a model was also tested by \citet{Alloin2007}, who used a truncated pseudo-isothermal profile to account for the common group halo. Here, we use the more realistic case of  including an NFW halo in addition to the six individual galaxies that make up the system. However, the position of the common halo and the mass-scale were poorly constrained by the data. Also, because we do not want to impose a bound state among these galaxies, we did not include this common halo as part of Model 2. Further observations at, for example, X-ray wavelengths may well constrain the position of a common halo, and reveal whether it is in a relaxed or disturbed dynamical state (e.g. \citealt{Fassnacht2008}).

The position angle of the external shear changes to $-61.2\pm2$~deg, which may suggest that any additional mass component causing this residual shear should be in the direction of G2 and G6 (see Fig. \ref{fig3}).  As a test, we included GA and GB (not spectroscopically confirmed as group members) in the model and found that there is a negligible change in the offsets between the observed and model predicted positions of the image-components, which is not surprising given the small contribution that they make to the total convergence. Indeed, by including these two galaxies the shear strength and its position angle do not change with respect to the Model 2 values. Therefore, we can exclude them as the  additonal mass component that may be responsible  for the residual external shear.

Finally, we note that although the structure of the background source is relatively unchanged when the group is included, the position of the background source does change to the extent that it no longer sits within the Einstein radius of G3 (Fig. \ref{fig5}). Therefore, without the additional convergence provided by the group of galaxies, MG~J0751+2716 would not be strongly gravitationally lensed.

\section{Discussion}
\label{disc}
In this section, we discuss our lens models for MG~J0751+2716 obtained from the VLBI imaging presented here.

\subsection{Precision lens modelling with global VLBI observations}

MG~J0751+2716 is one of the few quadruply imaged radio-loud gravitationally lensed quasars that show extended arcs on VLBI-scales. 
Our deep imaging detects the extended arcs at high significance, showing the complex surface brightness structure of the background source in unprecedented detail (Fig. \ref{fig1}). Never before have such extended (200--600~mas) gravitational arcs been detected at an angular resolution of a few mas. This detection allowed us to confidently identify lensed emission corresponding to the same source component, providing a very large number of constraints to the mass model that also sampled a large radial and tangential extent (Table \ref{tab1}). Moreover, these observations detect the faint counter-image B6 for the first time, providing a new additional constraint to the radial mass distribution of the lens. This component could be detected because of the excellent $\mu$Jy~beam$^{-1}$ sensitivity of the data, thanks to the large data recording rate (0.5 Gbit s$^{-1}$) and number of antennas used for this observation (Fig. \ref{fig2}). The advent of even larger recording rates (at 2--4~Gbit\,s$^{-1}$) and global VLBI arrays that contain over 25 antennas will routinely provide $\mu$Jy~beam$^{-1}$ surface-brightness sensitivities and excellent {\it uv}-coverage in a single synthesis observation. 

Using the constraints provided by the global VLBI imaging of MG~J0751+2716, we were able to infer the lens parameters with a high precision. By explicitly including the group of galaxies in the macro model, the uncertainties on the parameters are reduced significantly. For example, the lensing galaxy position is recovered with a precision of 0.6 and 0.1 per cent, respectively, for Models 1 and 2 (see Table \ref{tab_lensparameters}), which also corresponds to a factor of ten improvement in precision with respect to previous modelling that used MERLIN observations (50 mas FWHM beam size; rms 89~$\mu$Jy~beam$^{-1}$; \citealt{Lehar1997}). Moreover, our models constrain the Einstein radius with $\sim0.6$ per cent precision and the ellipticity at the order of 0.5 per cent (Table \ref{tab_lensparameters}). The slope of the mass density distribution $\gamma$ is found to be steeper than isothermal at the $4.2 \sigma$ level for Model 1 and at the $6.8 \sigma$ level for Model 2. The two main consequences of this precise lens modelling are explained in more detail below.  However, even if the lens parameters are recovered at sub-percent precision, they have significantly different values in Model 1 and Model 2. Therefore, we would like to highlight that the recovered parameter values are precise, but at least one, and possibly both, of the models are incorrect descriptions of the data, as we discuss further in Section 4.2. In other words, whilst the statistical uncertainties are quite small, the systematic uncertainties due to our model choices may be up to two orders of magnitude larger.

Our findings demonstrate that high resolution and high sensitivity observations are vital for testing complex mass models, as opposed to the standard assumption of a smooth power-law elliptical mass density distribution. However, such in-depth studies of the global mass distribution using mas-resolution observations have been mainly performed on compact (lensed) radio-cores, as a result of the selection criterium of most lensing surveys at radio wavelngths (e.g. JVAS/CLASS; \citealt{Myers2003, Browne2003}). For example,  using the same parametric lens modelling method applied here, the precision on the mass model parameters for CLASS~B0712+472  is of the order of $\sim10$~per cent when using the positions of the four compact lensed images measured with the VLBA at 1.7 GHz (10 mas FWHM beam size) as constraints \citep{Hsueh2017}. Also, by using the position of the four images of CLASS~B1555+375, as measured with MERLIN at 5 GHz (50 mas FWHM beam size), the precision of the mass model parameters is of the order of $\sim20$~per cent \citep{Hsueh2016}. 

As such, previous studies of the mass distributions of gravitational lenses have been limited by the number of systems that show extended structure, either due to the intrinsic source morphology or low brightness of any extended emission. To improve on this, we have started a high-sensitivity VLBI campaign of a carefully selected sample of gravitationally lensed radio sources with radio-bright Einstein rings or potentially extended arcs. The most promising sources have been followed-up with global VLBI imaging at mas resolution and high sensitivity, and will be presented in forthcoming papers. Nevertheless, the next generation of interferometers (i.e. Square Kilometer Array; SKA) will allow the discovery of $\sim 10^5$ gravitational lenses with both compact and extended structure, increasing by several orders of magnitude the number of systems suitable for testing mass distributions on mas-scales \citep{KoopmansSKA2004, McKeanSKA2015}. In addition, the next generation of optical/infrared telescopes will have a sensitivity and angular resolution that is comparable to global VLBI observations, for example the European-Extremely Large Telescope (E-ELT), which will allow a multi-wavelength test of lensing macro models at radio and infrared wavelengths.

\subsection{Evidence for additional mass structure}
\label{astrometric_anomaly}

In Fig. \ref{fig6} we show the image position residuals for Model 1 and Model 2. We find that the image position residuals do not correlate with any particular group of lensed images, but they are scattered almost uniformly in Right Ascension and Declination.
A possible explanation for the offset between the observed and model-predicted positions is the presence of some additional mass structure within the lensing galaxy or along the line-of-sight that has not yet been taken into account in the lens models presented here.  

Based on the methodology provided by \citet{Despali2017}, we find that the combination of source and lens redshifts for MG~J0751+2716 leads to a projected number density of low-mass ($M\sim10^6$ M$_{\odot}$) line-of-sight haloes per arcsec$^2$ of the order of a few tens for a CDM scenario.
This additional mass can be contained, for example, in sub-haloes associated with the G3 dark matter halo, which can change the deflection angle and, therefore, shift the position of the lensed images from what is expected from a smooth mass distribution \citep{Wamb1992, MetcalfMadau2001, Metcalf2002, Inoue2003, Inoue2005a, Inoue2005b, Sluse2012}. This method has been used to quantify the level of substructures in the intermediate mass regime ($\sim 10^{8}$~M$_{\odot}$) from adaptive optics imaging of extended gravitational arcs \citep{Vegetti2012} and from spectro-imaging of the narrow-line region of lensed quasars \citep{Nierenberg2014}. However, only VLBI observations can directly resolve the small-scale astrometric shifts due to very low mass haloes ($\sim 10^{6}$~M$_{\odot}$; \citealt{McKeanSKA2015}) because of the excellent astrometric information provided by the data  \citep{Chen2007,KeetonMoustakas2009}. 

For example, VLBI observations of CLASS~B0128+437 at mas resolution revealed astrometric offsets of between 5 to 10 mas (much larger than the intrinsic astrometric precision) that have been ascribed to the presence of substructure within the main lensing galaxy \citep{Biggs2004}. Also, multi-frequency global VLBI observations of MG~B2016+112 confirmed that the astrometric anomaly observed for this system could be entirely attributed to a luminous satellite associated with the lensing galaxy \citep{Koopmans2002, More2009}. In the case of MG J0414+0534, the high resolution imaging from VLBI was used to infer the position and mass of a dark sub-halo \citep{MacLeod2013}.  Moreover, the radio-loud lensing systems CLASS~B1933+503, CLASS~B1555+375 and CLASS~B0712+472 have extended arcs, which also show hints of a disturbed morphology or anomalous flux-ratios at radio wavelengths \citep{Jackson1998, Marlow1999, Norbury2001}. Subsequently, optical imaging of theses systems revealed that the lensing galaxy is a late-type galaxy and by adding the disk as the additional mass component the position of the lensed images could be completely recovered \citep{Suyu2012, Hsueh2016, Hsueh2017}.

Nevertheless, given the complexity of the mass model for this lensing system, we cannot draw stringent conclusions on the origin of the observed astrometric anomaly. Only a Bayesian grid-based analysis that takes into account the flux density distribution of the entire lensed arcs and performs lens-potential corrections directly in the visibility plane can test whether our parametric models are too simplistic for this system, or if there is the need for extra-mass in the model.

\subsection{Evidence in favour of the two-phase galaxy formation scenario}
\label{density_profile}

Both Model 1 and Model 2 find that the mass density profile for G3 is steeper than isothermal with $\gamma_1 = 2.08 \pm 0.02$ ($4.2 \sigma$ level) and $\gamma_2 = 2.16 \pm 0.02$ ($6.8 \sigma$ level), at the Einstein radius $b_1 = 0.4025 \pm 0.0008$ and $b_2 = 0.307 \pm 0.002$~arcsec, respectively.  
These density slopes are consistent within 1$\sigma$ of the distribution of slopes from the Sloan Lens ACS Survey (SLACS) sample of lenses \citep{Auger2010}, which on average is $\gamma= 2.078 \pm 0.027$, therefore mildly steeper than isothermal \citep{ Koopmans2009, Auger2010, Barnabe2011, Sonnenfeld2013}. Therefore, the MG~J0751+2716 lensing galaxy has a similar density profile when compared to other early-type galaxies that act as strong  gravitational lenses at a similar redshift.

Nevertheless, this density profile slope can be considered as evidence for the so-called \textsl{two-phase} galaxy formation scenario for G3, which is a low-mass early-type galaxy (with velocity dispersion $\sigma_{v} = 101$~km\,s$^{-1}$; \citealt{Alloin2007}) in a very rich environment (Fig. \ref{fig3}, but also \citealt{Tonry1998, Momcheva2006, Alloin2007, Momcheva2015, Wilson2016}).  It has been shown that early-type satellite gravitational lenses within groups or clusters of galaxies (as in the case of G3) are better modelled with mass density profiles  that are steeper than isothermal \citep{Rusin2002, Dobke2007, Auger2008sbs1520}.  Also, the SLACS lenses show a super-isothermal mass density profile when associated with a perturbing companion galaxy (in this case it is likely G1), suggesting that this steepening can be attributed to their possible interaction  \citep{Auger2008}. Moreover, from observations and simulations it was found that low-mass (and compact) early-type galaxies have a mass slope $\gamma$ steeper than isothermal, while high-mass early type galaxies have a shallower $\gamma$ \citep{Barnabe2011, Sonnenfeld2012, Remus2013, Dutton2014, Tortora2014}. 

The theoretical scenario for explaining $\gamma > 2$ in low-mass non-isolated galaxies (called \textit{two-phase} scenario) regards the interaction with their companion galaxies, in particular, their merger history \citep{Guo2008, Oser2010, Johansson2012, Remus2013, Dubois2013}. In this framework, at the early stages of galaxy formation gas-rich mergers lead to an enhanced \textsl{in situ} star formation; the dissipative process due to \textsl{in situ} star formation cause a mass density profile that is super-isothermal with an increased baryonic matter content in the central regions of the  galaxy. Then, after $z\sim 2$, the merging events lead to a reordering of the early-type galaxy into an isothermal state, because the dissipative processes are not dominant anymore, and the galaxy growth is principally driven by gas-poor mergers. The data in hand for MG J0751+2716 are consistent with this model, given the dense environment of the system and a density profile that is steeper than the isothermal case.

\section{Conclusions}
\label{conc}

We have presented sensitive global VLBI observations at 1.65 GHz of the radio-loud quasar MG~J0751+2716, which is gravitationally lensed by a foreground group of galaxies to produce gravitational arcs that are extended by 200 to 600~mas; these data represent the highest angular resolution imaging of extended gravitational arcs from a gravitational lens. Our observations demonstrate that mas resolution observations of gravitationally lensed radio-sources can provide a large number of constraints to the lensing mass model, which can be used to search for any deviation from a globally smooth mass distribution. By using the positions of four quadruply-imaged components and two doubly-imaged components, we investigate a simple single-lens model and a more realistic mass model for the group of galaxies that is associated with the main lensing galaxy. We find that from these constraints, we are able to infer the lens model parameters with a precision of less than a percent, even though our models are not accurate enough to fit the positions of the observed images to the measurement error level. Furthermore, both models suggest an inner density slope for the main lensing galaxy that is steeper than isothermal. This is consistent with studies of other low-mass early-type satellite galaxies in dense environments, and is in agreement with the \textsl{two-phase} galaxy formation scenario. This is important, because more than 50 percent of galaxies are found to lie in groups, at least locally, and to date, there is not a complete picture of the total projected mass distribution of galaxy groups.

Due to the excellent sensitivity and high angular resolution of the VLBI imaging, we find there is a discrepancy between the observed and predicted positions of the lensed images, with an average position rms of the order of 3 mas for the simple parametric models tested here. At this stage, it is not clear if these deviations are due to some additional mass structure in the form of a population of low mass sub-haloes that are either part of the lensing group or along the line-of-sight, or if the complexity of the group environment is not being fully taken into account by the parametric models. In a future paper, we will present modelling with a grid-based source surface brightness distribution,  which is fitted directly with the visibility data, that will allow the complete set of extended gravitational arcs to be used as constraints. In addition, with the improved source model, we will be able to test non-parametric lens models using grid-based corrections to the gravitational potential, based on the methodology of \citet{Vegetti2009}, which will shed light on the cause of the astrometric anomaly seen in the compact lensed components of MG~J0751+2716.


\section*{Acknowledgments}
LVEK is supported through an NWO-VICI grant (project number 639.043.308). CDF acknowledges support from the U.S. National Science Foundation, via grant AST-1715611. This project has received funding from the European Research Council under the European Union's Horizon 2020 research and innovative programme (grant agreement No 758853). DJL acknowledges support from the European Research Council (ERC) starting grant 336736-CALENDS. The National Radio Astronomy Observatory is a facility of the National Science Foundation operated under cooperative agreement by Associated Universities, Inc. The European VLBI Network is a joint facility of European, Chinese, South African and other radio astronomy institutes funded by their national research councils.  Scientific results from data presented in this publication are derived from the following EVN project code: GM070. Based on observations made with the NASA/ESA Hubble Space Telescope, obtained from the Data Archive at the Space Telescope Science Institute, which is operated by the Association of Universities for Research in Astronomy, Inc., under NASA contract NAS 5-26555. These observations are associated with program 7495.

\bibliographystyle{mnras}
\bibliography{SHARP-V}

\bsp	
\label{lastpage}
\end{document}